\documentclass{ws-ijmpa}
\usepackage{amssymb}
\usepackage{bm}

\newcommand{\be}{\begin{equation}}
\newcommand{\ee}{\end{equation}}
\newcommand{\bea}{\begin{eqnarray}}
\newcommand{\eea}{\end{eqnarray}}
\newcommand{\hf} {\frac{1}{2}}

\def\journal#1#2#3#4{{#1} {\bf #2}, #3 (#4)}

\def\eq#1{(\ref{#1})}

\def\ord#1{{\cal O}(#1)}

\def\mr#1{{\mathrm{#1}}}
\def\Tr{\mr{Tr}}
\def\hi{\frac1\hbar}

\begin{document}

\markboth{V. Pangon, S. Nagy,J. Polonyi, K. Sailer}
{Onset of symmetry breaking by the functional RG method}

%
\catchline{}{}{}{}{}
%

\title{Onset of symmetry breaking by the functional RG method}

\author{V. Pangon}

\address{Gesellschaft f\"ur Schwerionenforschung mbH, Planckstr. 1, D-64291 Darmstadt, Germany}
\address{Frankfurt Institute for Advanced Studies, Universit\"at Frankfurt, D-60438 Frankfurt am Main, Germany\\
V.Pangon@GSI.de}

\author{S. Nagy}
\address{Department of Theoretical Physics, University of Debrecen, Debrecen, Hungary\\
nagys@dtp.atomki.hu}

\author{J. Polonyi}
\address{Strasbourg University, CNRS-IPHC, BP28 67037 Strasbourg Cedex 2, France\\
polonyi@IReS.in2p3.fr}

\author{K. Sailer}
\address{Department of Theoretical Physics, University of Debrecen,
Debrecen, Hungary\\
sailer@phys.unideb.hu}

\maketitle

\begin{history}
\received{Day Month Year}
\revised{Day Month Year}
\end{history}

\begin{abstract}
A numerical algorithm is used to solve the bare and the effective potential for the scalar
$\phi^4$ model in the local potential approximation. An approximate dynamical Maxwell-cut 
is found which reveals itself in the degeneracy of the action for modes at some scale. This
result indicates that the potential develop singular field dependence as far as one can see
it by an algorithm of limited numerical accuracy.

\end{abstract}

\ccode{PACS numbers: 05.10.Cc,11.30.Qc,64.60.ae}

\section{Introduction}
The functional renormalization group method \cite{bagn,berg,pol} is a promising tool
to study and solve strongly coupled quantum systems. It is based on the
gradual elimination of degrees of freedom by decreasing a continuous parameter
of the dynamics, traditionally chosen to be the cutoff, $k\to k-\Delta k$.
The novelty is the appearance of a new small parameter, the relative
change of the cutoff, $\Delta k/k$, which simplifies enormously
the renormalization group equation. 

One assumes at the derivation of the Wegner-Houghton \cite{wh} or the 
Polchinski \cite{polch} equations the availability of the loop or the perturbation expansion, 
respectively \cite{pol}. The two schemes usually agree, except in the 
presence of non-trivial saddle points.
But the true expansion parameter for the change of the Wilsonian action is the 
naive small parameter multiplied by $\Delta k/k$ and this latter factor renders 
the leading order expressions for the evolution equation exact in the differential 
equation limit, $\Delta k\to0$ \cite{pol}. These equations are based on an approximate 
evaluation of the blocking transformation and their solution resums the expansion series.

There is a way to establish the renormalized trajectory by formal manipulations,
without computing any path integral during the evolution. One introduces a control
parameter, the cutoff \cite{nicoll,wetterich,morris} or any other continuous
parameter of the action \cite{alexandre} which is chosen in such a manner that
its evolution guides the theory from a perturbative initial condition to the physical one 
in a smooth, differentiable manner. The evolution equation is written formally only, in terms of 
a generator functional without carrying out functional integration. It is natural to use 
the effective action as the generator functional for the one particle irreducible vertex 
function for this role because it is assumed to be local which makes the evolution equation
easier to truncate. The leading order term of the evolution equation in $\Delta k$ gives 
a functional differential equation whose validity is not based on any expansion scheme 
of the physical theory. The only functional integration to carry out is in the initial 
condition where the assumed perturbative effective action is constructed. Similar evolution 
equation can be obtained by the average action method \cite{ringwald,tetradis,wett} but this is 
based on the bare action and assumes the availability of the loop expansion.

The only unavoidable approximation of the functional renormalization group method
is the projection of the exact renormalization group equation into a functional space for the 
action where the solution is easier to find. This step, the choice of ansatz for the action, is 
guided by our intuition and our knowledge about the dynamics and provides a flexible 
truncation scheme.

This scheme has two serious drawbacks. One is its weakness of handling degrees 
of freedom whose dynamics, the dependence of the action on their coordinate, is weak. 
In fact, the methods tracing the evolution of the Wilsonian action may fail 
because the loop or the perturbation expansions are not reliable for a degenerate action.
The functional Legendre transformation, employed in the derivation of the
effective action becomes singular when it reaches the convexity limit and renders the effective action an unreliable
tool for the study of an almost degenerate system. Note that the case of 
the degenerate action is more complicated than those of the usual strongly coupled situation because
the integrand of the path integral is constant and all field configurations
contribute with equal strength. Such a degeneracy  poses a danger for any schemes,
based either on the bare or the effective action because it renders the truncation
of the functionals used far more difficult. 

There is yet another problem related to such a high degree of degeneracy. Like symmetry 
breaking, it appears in a discontinuous manner as the control parameter is changed. The 
physical system is approached along an artificial line of theories where we integrate 
the evolution equation. When the evolution equation develops nonintegrable singularity
when the degeneracy is reached then the method fails.

The prototype of degenerate systems is a condensate or the spontaneous breakdown of 
a symmetry. The mixed phase, observed for such a system for certain values of the 
order parameter, is based on the competition of two different orders
and the position of the domain wall, separating such realizations represents a 
degenerate collective mode. When equilibrium concepts are applicable 
like for Euclidean Quantum Field Theory, then the Maxwell-cut renders 
the action strictly degenerate in some collective coordinate. The degeneracy
opens the possibility that completely new degrees of freedom appear, such as
the gas molecules in the liquid-vapor transition. The goal of this work is to 
investigate this problem in its simplest realization, in the framework of the single
component, scalar $\phi^4$ field theory in four dimensional Euclidean space-time 
by means of the reconstruction of the renormalized trajectory both for the Wilsonian 
action and for the effective action.

There have been several papers addressing the symmetry broken phase within the
framework of the functional renormalization group scheme. The evolution 
of the running mass and quartic coupling strength was followed in the
$O(N)$ symmetric scalar model\cite{ringwald} and instability, 
indicated by the appearance of saddle point contributions to 
the blocking relation, was found. But instabilities, more precisely eventual
singularities of the flow may render the strategy, based on the integration of the 
evolution equation inapplicable as explained above. Further problems are expected
when smooth cutoff is applied. The saddle point was believed to render
the action is non-degenerate in the unstable region depends on the arbitrarely 
chosen artificial smooth IR cutoff. This suggests that the original dynamics without the 
hand made suppression is strongly coupled and a polynomial truncation of the potential 
can not be justified. Moreover, the original plan of removing the smooth cutoff from the 
theory without trace is realized only when the cutoff can be lowered down to the IR end 
point in a consistent manner which does not seem to be the case here.

The high degree of degeneracy requires unusual care in truncating the evolution
equation. It would seem natural to use $\epsilon(k)$, the variation of the action
at the cutoff scale $k$, as a small parameter and ignore terms $\ord{\epsilon}$
in the evolutions equation as the degeneracy is approached \cite{tetradis}. The 
restriction to quartic potential can be avoided in this manner and the local potential 
approximation to the action produces a scale independent, universal parabolic 
potential for small enough field values, in the degenerate regime, interpreted
as a fixed point. But the omitted, $\ord{\epsilon}$ terms are actually
playing an important role. In fact, the small parameter of the loop expansion supporting 
the average action scheme is proportional to $1/\epsilon$ and the the error of such a
truncated "renormalization group improved" solution is $\ord{\epsilon^0}$. For instance,
the $\ord{\epsilon}$ terms regulate the otherwise ill-defined 0/0 expressions in the 
$\beta$-function as the degeneracy is approached. This issue, the potential
singularities of the flow induced by the $\ord{\epsilon}$ terms is addressed in the 
present work by means of a more reliable numerical algorithm.

The next order in the gradient expansion,
the wave function renormalization constant was included in a similar study
\cite{wett} but the local potential was again truncated to quartic order.
A formal proof of regularity of the evolution equation for the effective action
is presented in Ref. \cite{lipa} for smooth cutoffs with sufficiently strong 
momentum dependence. This restriction and the assumed analytical dependence of the
evolution equation on the field indicates that the analytical continuation explored
in this work is not unique. There is no such a problem when the evolution of the
bare action is followed \cite{wh} because the modes are eliminated by taking into account
their physical coupling to the rest of the system. The smooth cutoff procedure, an
unavoidable step within the gradient expansion ansatz for the effective action,
introduces artificial dynamics for the modes partially eliminated. The dependence on this
arbitrary modification of the dynamics is canceled when the evolution equation is 
integrated down to the ultimate IR end point exactly. But any truncation or singularity 
which prevents us to reach the the IR end point renders the cancellation questionable 
and the scheme dependence spoils the results.

The new element of our study is a sufficiently powerful numerical algorithm to solve the evolution
equation for a general potential without assuming power series form. Such a generalization
raises the hope of finding out what happens with the two potential obstacles of the
evolution equation method mentioned above. We mention that this numerical method
was needed for the more satisfactory description of the symmetry broken symmetric phase of 
the sine-Gordon model \cite{sg,vincent}. Though one can not reach a final answer with
numerical method of limited accuracy it seems to be likely that a strong degeneracy
and singularities are established at finite scale.

One should mention that the integration of the evolution equation without assuming 
a polynomial ansatz for the local potential has been considered in a number of works\cite{adams,schaefer,bohr}
where number of points chosen along the field axis to monitor the potential was in the 
order of 60-100. In view of the results presented below, namely that the 
increasing degeneracy of the action as the evolution approaches a finite scale makes the 
evolution equation stiff and requires a numerical precision which can not be achieved even 
by using spline functions on 10 000 collocation points the results reported in these publications are not reliable.

The organization of this paper is the following. The evolution equation, its
scaling regimes and its validity is discussed briefly in Section \ref{evoleq}.
In Section \ref{condeb} we discuss the ways condensate and symmetry breaking manifest
themselves in the evolution equation method. The numerical solution of the evolution
equations is presented in Section \ref{numsol}. Finally, Section \ref{sum} contains the 
summary of the results. \ref{algorithm} is provided to give some details of 
the numerical algorithm used. Finally, \ref{wilsfis} contains the check of
applied numerical algorithm against the Wilson-Fisher fixed point.

\section{Evolution equations}\label{evoleq}
The Euclidean partition function of the model, investigated in this paper 
is given by the path integral
\be\label{pint}
Z=\int D[\phi]e^{-\hi S[\phi]},
\ee
which is rendered finite by a large but finite sharp momentum space cutoff $\Lambda$. 
The evolution of the bare
action will be followed by lowering the sharp UV cutoff and the effective action will be 
constructed by lowering an additional smooth IR cutoff. We shall restrict our attention to 
the local potential ansatz in the rest of the paper where the bare or the effective action
is written as a usual kinetic term plus the space-time integral of a local potential $V(\phi)$.

\subsection{Bare potential}
We lower the cutoff, $k\to k-\Delta k$, by splitting the field variable as
$\phi=\phi'+\tilde\phi$ in such a manner that the supports of the Fourier 
transforms of $\phi'$ and $\tilde\phi$ are $|p|<k-\Delta k$ and $k-\Delta k<|p|<k$,
respectively, and integrate over the UV field,
\be\label{blocking}
e^{-\hi S_{k-\Delta k}[\phi']}=\int D[\tilde\phi]e^{-\hi S_k[\phi'+\tilde\phi]}.
\ee
We assume that the loop expansion is applicable, yielding
\be\label{whiter}
S_k[\phi'+\tilde\phi_{cl}]-S_{k-\Delta k}[\phi']=-\frac\hbar2\Tr\ln\frac{\delta^2S_k[\phi'+\tilde\phi_{cl}]}
{\delta\phi'\delta\phi'}+\ord{\hbar^2},
\ee
where $\tilde\phi_{cl}$ denotes the saddle point in the space of the UV field configurations
and $\ord{\hbar^2}$ stands for the higher-loop contributions. The action is assumed to take the form
\be\label{ansatz}
S_k[\phi]=\int d^4x\left[\hf(\partial\phi)^2+V_k(\phi)\right]
\ee
within the local potential approximation, followed in this work. Since there is
a single potential to keep track thus the choice of a homogeneous field, $\phi'_x=\Phi$ is natural. 
We evaluate the trace in the Fourier space and assume the vanishing
of the saddle point $\tilde\phi_{cl}=0$. The result is the Wegner-Houghton equation \cite{wh},
\be\label{wh}
\partial_kV_k(\Phi)=-\frac{\hbar k^3}{16\pi^2}\ln[k^2+V_k''(\Phi)],
\ee
in the limit $\Delta k\to0$ where $V''(\phi)=\partial^2_\phi V(\phi)$. The 
higher loop contributions are suppressed by $\Delta k/k$ and are negligible
in the differential equation limit. When the equation is integrated numerically then each
step $k\to k-\Delta k$ generates one further order in the loop expansion and
the solution of the differential equation, obtained in the limit $\Delta k\to0$,
contains the contributions of all connected Green functions at vanishing momentum.

We turn now to the characterization of the different scaling regimes we may encounter.
We expect two long and distinct scaling regimes, and UV and and IR one separated by a
crossover scale $k_{cr}$. We shall be interested in the low energy content of the theories 
and shall follow the evolution towards the IR direction and address no UV issues, such as 
triviality. The qualitative behavior of the running coupling constants $g_n=\partial^n_\Phi V(0)$ 
can easily be reconstructed. The corresponding beta functions, 
$\beta_n=k\partial_kg_n$, are obtained by taking $n$ successive derivatives of Eq. \eq{wh} 
with respect to $\Phi$. The result is a sum, consisting of terms like 
\be\label{betfccontr}
(-1)^{N_n}c_n\frac{\prod_{j=3}^ng_j^{\ell^{(n)}_j}}{(k^2+g_2)^{N_n}},
\ee
where $c_n$ is a dimensionless, positive number, 
$\sum_j\ell^{(n)}_j=N_n$ and $\sum_j(4-j)\ell^{(n)}_j=4-n$ \cite{pol}.
The dominant term of the sum, if there is one in an asymptotic scaling regime, is selected by the 
competition between the numerators and denominators. The alternating sign strongly reduces
the size of the asymptotic scaling regimes. Consider first the most important beta function,
\be
\beta_2=-\frac{\hbar k^4}{16\pi^2}\left[\frac{g_4}{k^2+g_2}-\frac{g_3^2}{(k^2+g_2)^2}\right].
\ee
It makes the mass square, $g_2$, increasing during the evolution in a $\phi^4$ theory 
as long as $k^2+g_2>0$. In the asymptotic UV scaling regime the denominator in 
Eq. \eq{betfccontr} is made large by the cutoff and the power of the propagator must be 
minimized to find the dominant term, giving $\beta_n\sim g_n/(k^2+g_2)$. 
The asymptotic beta function of the almost degenerate theory starting at $k_{cr}$
where $k^2+g_2$ can be considered small is dominated by the term with the maximal power of 
the propagator, $\beta_n\sim (g_4/k^2)^{n/2-2}$. 

We shall be interested in what happens in the vicinity of $k_{cr}$ and it is instructive 
to follow first a naive argument as to what happens around $k_{cr}$, by ignoring the 
loop-generated evolution beside the tree-level contributions\cite{jean}. The coupling are 
constant when the loop-generated evolution is ignored and the inverse propagator crosses 
zero at $k_{cr}^2=-g_2$ due to the decrease of the kinetic energy. The inverse propagator, 
the argument in the logarithm of Eq. \eq{wh}, is an eigenvalue of the second functional derivative 
of the action. When a negative eigenvalue appears the integral of Eq. \eq{blocking} has a
non-trivial saddle point. Therefore, if the inverse propagator changes sign then
the renormalization group equation has tree-level contribution. The tree-level contributions 
of plane waves, alias domain walls, to the evolution equation lead to the Maxwell-cut,
the degeneracy of the IR local potential, $V_{k=0}(\Phi)$, between the stable vacua, 
$|\Phi|<\Phi_\mr{vac}$. As a result, the action will be degenerate for an increasing interval 
of $\Phi$ around zero as $k$ is further decreased. The action $S_k[\phi]$ is 
therefore degenerate for modes $p\approx k$ which are best described by such an 
effective theory. We arrive at a dynamical extension of the Maxwell-cut in this 
approximate solution. The singularity of the tree-level flow renders the stategy of
integrating the evolution equation unreliable as explained in the Introduction.
Therefore we do not pursue this scenario and turn to the main goal of this paper, the 
verification if the functional RG method remains valid when loop-corrections are taken into account.

We finally address the question of the applicability of the loop expansion in the functional 
integral \eq{blocking}, assumed in deriving the evolution equation \eq{wh}. 
The dimensionless small parameter of the loop expansion is 
$\epsilon_k(\Phi)=\hbar\Delta k\ln[1+g_{2k}(\Phi)/k^2]/k$.  By setting
the book-keeping variable to unity, $\hbar=1$, there are two ways this
parameter may become large: either we reach the IR end point, $\Delta k/k\to1$ or
the logarithm function explodes because the action is approximately degenerate. 
The former case requires to stop the integration of the evolution 
equation \eq{wh} slightly before $k=0$. The results obtained in this manner are useful because
they characterize the infrared end point, $k=0$ if the theory has a convergent thermodynamical 
limit, an assumption one is usually ready to agree with. Thus the real danger for the validity 
of the evolution equation comes from the second mechanism, the approximate degeneracy of the action 
occurring already at finite scale.

\subsection{Effective potential}
The effective action is obtained by adding a quadratic expression to the action
\be
S[\phi]\to S[\phi]+\hf\int d^4x\phi R_k(-\Box)\phi
\ee
which suppresses the modes below the scale $k$. For $k\approx\Lambda$ the fluctuations
are weak and the theory is supposed to be perturbative and for $k=0$ the artificial
term is chosen to be vanishing. The evolution equation for the local potential of an
ansatz \eq{ansatz} for the effective action can easily be derived \cite{nicoll,wetterich,morris},
\be\label{smooth}
\partial_kV_k(\Phi)=\hf\int\frac{d^4k}{(2\pi)^4}\frac{\partial_kR_k(p^2)}{p^2+V''_k(\Phi)+R_k(p^2)}
\ee
where $\hbar=1$ was set as in the rest of this work. We shall use power functions 
\be\label{regpow}
R_k(p^2)=ap^2\left(\frac{p^2}{k^2}\right)^{-b}
\ee
with $a,b>0$, which realize a smooth IR cutoff for simplicity. The beta functions are 
more involved as in the case of sharp cutoff but one expects no qualitative differences.
The effective action must be convex for any momentum component of the field. In particular,
for a single Fourier component with infinitesimal amplitude we have the bound
\be
0< min_{p}\Big\{p^2+R_k(p^2)\Big\}+V''_k(\Phi)=\overline{k}^2+V''_k(\Phi)
\ee
which imposes convexity on the true effective potential for $k=p=0$. For the 
choice of regulator of Eq. \eq{regpow} $\overline{k}^2$ reads
\be\label{meas}
\overline{k}^2=b(b-1)^{\frac{1-b}{b}}a^{\frac{1}{b}}k^2=\tilde{k}^2k^2
\ee

\section{Condensate and symmetry breaking}\label{condeb}
We make a closer view in this section on the way the condensate and the symmetry breaking 
is handled in the two schemes introduced above. Let us first point out that any perturbative 
method is in general local in the internal space, in the space of the field amplitude. 
In fact, the very essence of perturbation expansion is the assumption that the field 
fluctuates weakly around some mean value and explores the dynamics, characterized by some 
action in its vicinity only. On the other hand, the genuine nonperturbative schemes
must take into account the interference among states with largely differing field
expectation values. This rather general feature is important when the action is
strongly degenerate.

The numerical simulations of the single component $\phi^4$ model
in lattice regularization show that the zero momentum component of the field
fluctuates around a nonvanishing value in the symmetry broken vacuum. This seems to be
consistent with perturbation expansion which predicts a slightly modified Gaussian 
wave functional at the vacuum expectation value. In the same time the effective potential 
which is supposed to be proportional to the logarithm of the probability distribution of 
this field component is constant in between the vacuum expectation values due to the 
Maxwell-cut. What happens is that large amplitude fluctuations render the perturbative
picture invalid inside the mixed phase where effective action is degenerate. These
large amplitude fluctuations are supposed to be $\hbar$-independent, inhomogeneous
semiclassical modes. Being inhomogeneous, the integration over their zero modes, 
arising from the breakdown of external space-time symmetries restore the homogeneity
of the vacuum. Thus the large amplitude fluctuations explore the action far from the
vacuum configuration and the corresponding equations are nonlocal in the internal 
space. Therefore, if a nonperturbative method provides us a ``solution`` of the theory 
it must find the potential and the other ingredients of the action for all possible value 
of the field. A nonperturbative solution of the symmetry broken vacuum  must explain the 
presence of the Gaussian looking peak in the probability distribution of the zero momentum 
component of the field despite the degeneracy of the action at one side of the vacuum. 

Another lesson is the need of a more careful definition of condensate. Guided 
by the analogy with Bose-Einstein condensation we call a state condensate if its
occupation and structure is independent of $\hbar$. The integration over the zero modes
may lead to a decohered superposition of states with macroscopically different
field expectation values without having nontrivial order parameter expectation value and 
symmetry breaking.

Let us now focus on the functional renormalization schemes from the point of view of the 
two limiting factors, the issue of nonintegrable singularities and the large degree of 
degeneracy. The simplest realization where the control parameter is momentum independent 
and quadratic in the field is the Callan-Symanzik scheme \cite{alexandre}. The evolution 
of the bare mass square from a sufficiently large, positive initial condition to its 
physical value generates a path in the space of theories which interpolates between a 
perturbative and the true theory. If the initial and the final points are in different 
phase then the trajectory passes singular point and the integrability of the evolution 
equation can not be assured by our present analytical means. We need a different, better 
chosen path between calculable initial condition and a fixed physical theory for the 
evolution method be useful. A high degree of degeneracy
is customary in field theory but it used to appear at the IR fixed point. The
novel, more violent feature of the singularity discussed in this work is that it
arises at finite scale. The problems come not only from the fact that 
the phase space available for the soft modes is larger than at the IR fixed point
but that we should traverse the singularity by integrating the evolution equation.

In the evolution of the bare action the propagator of the Euclidean theory, the inverse 
of the argument of the logarithm function in Eq. \eq{wh} remains positive all the way 
during the evolution in the symmetrical phase. The emergence of a degeneracy is signalled 
by reaching a zero for the inverse propagator at some value of the sharp momentum space 
cutoff $k=k_{cr}$ first at $\Phi=0$. If this happens then the further lowering of the cutoff
generates a shallow, $\ord{\delta k}$ minimum at $\Phi\ne0$ which leads to 
inhomogeneous condensate \cite{jean}. As discussed above, the accuracy of this solution 
is lowered to $\ord{\hbar}$ and we see the semiclassical vacuum structure emerging. Such
a vacuum consists of the gas of domain walls and the integration of the zero modes
due to their broken space-time symmetries restores the homogeneity of the vacuum.

It is important to keep in mind that during the successive integration we have carried out
during the lowering of the cutoff to the running scale $k$ we eliminate infinitely many variables.
Therefore this path integral, what we shall call an incomplete theory, is without 
homogeneous order parameter for $k\ne0$, but still may possess nontrivial phase structure. 
The incomplete theory is in the symmetrical phase in the asymptotic UV scaling regime but 
must arrive at the phase boundary at some scale $k_{tr}$ if the complete theory has 
symmetry broken vacuum. If condensate is encountered in the incomplete theory then it
should not show up later than reaching the phase boundary, $0\le k_{tr}\le k_{cr}$.
The Wegner-Houghton equation faces the problem related to the degeneracy and the
nonintegrability at $k=k_{cr}$ and $k=k_{tr}$, respectively.
Despite this difficulty on expects that the problem
of nonintegrablity is less serious than the other. The reason is that the renormalized
trajectory of a local theory is always continuous \cite{israel,enter}. This general
theorem is realized in this scheme in a simple manner, a loop integral within an 
infinitesimal volume in the momentum space, the functional trace on the right hand side  
of Eq. \eq{whiter} can not produce discontinuity in Eq. \eq{wh} as long as the integrand 
is regular or its singularity is integrable. The regularity of the integrand is guaranteed
for nonvanishing argument of the logarithm function, suggesting $k_{tr}=k_{cr}$, 
an indication that the problems of the degeneracy and the nonintegrability are intermingled.
The right hand side of Eq. \eq{whiter} remains always finite due to the logarithm function
and the bare theory displays finite beta functions. Thus the question we intend to
clarify below is the eventual loss of the Wegner-Houghton equation due to the
emergence of a high degree of degeneracy and a condensate at nonvanishing scale, $k_{cr}>0$.

The path integral implied in the evolving effective action extends over the complete
space of field configurations and it is straightforward to decide if the incomplete theory
which is dominated by modes in the scale interval $[k,\Lambda]$ is in the symmetry broken 
phase by inspecting the expectation value of the order parameter, the
homogeneous component of the field. One expects again two characteristic scales satisfying
the same inequality as before. The issue of the possible nonintegrability of the 
evolution equation at $k_{tr}$ is less clear than in the case of sharp
cutoff because the functional integration corresponding to an infinitesimal blocking
$k\to k-\Delta k$ always extends to the whole configuration space and it is difficult to
assess its dominance by some smaller subspace. In addition, the problem seen in the case 
of the Callan-Symanzik scheme is more relevant, too. In fact, our incomplete theory
includes all IR modes and when it passes the nonperturbative critical theory singularities 
arise in the $k$ dependence unless the suppression in the deep IR region is
strong enough. Due to lack of the logarithm function on the right hand side of Eq. 
\eq{smooth} the renormalized trajectory may develop singular beta functions.

Our strategy, followed below, is to monitor the degeneracy $\overline{k}^2+V''_k(\Phi)$,
cf. Eq. \eq{meas} for the two renormalization group schemes with the goal to find out the actual value of $k_{cr}$.

\section{Numerical solution}\label{numsol}
In order to avoid the need of the polynomial ansatz for the potential which can be
justified by means of the perturbation expansion only, we seek the solution of the
evolution equations for a general function $V_k(\Phi)$,
represented by spline polynomials. Thus this equation is used to generate the 
evolution of the coefficients of the spline functions, see Appendix \ref{algorithm} for details.
The fundamental difference between the expansion and the spline methods is the
handling of boundary conditions. In fact, this is an implicit assumption for
the expansion method but requires explicit equation for the spline representation.
We have to provide two functions to make the solution of Eq. \eq{wh} unique.
The range of the field variable will be restricted to the interval $(0,\Phi_\mr{max})$
where $\Phi_\mr{max}$ is sufficiently large. One of the boundary conditions, $V'_k(0)=0$, 
expresses the $Z_2$ symmetry of the theory. The other condition of the form
$0=\alpha V_k(\Phi_\mr{max})+\beta V'_k(\Phi_\mr{max})$ will be used with the trivial
choice, $\alpha=\beta=0$, in the algorithm which remains well defined in this case. 
The error induced by this null-boundary conditions is estimated by considering the solution 
in larger interval, by studying the convergence of the solution in the limit 
$\Phi_\mr{max}\to\infty$. It was carefully checked that this limit indeed leaves behind a 
stable, convergent numerical solution. This numerical approach was tested in 3 dimensions
and Appendix \ref{wilsfis} contains a brief description of the recovery of the 
Wilson-Fisher fixed point. We used a spline function splitting the resolution interval 
into 20000 subintervals, each of them carrying a degree 3 Chebyshev polynom, to map 
out the field dependence.

\subsection{Scaling laws}
\begin{figure}
\centerline{\psfig{file=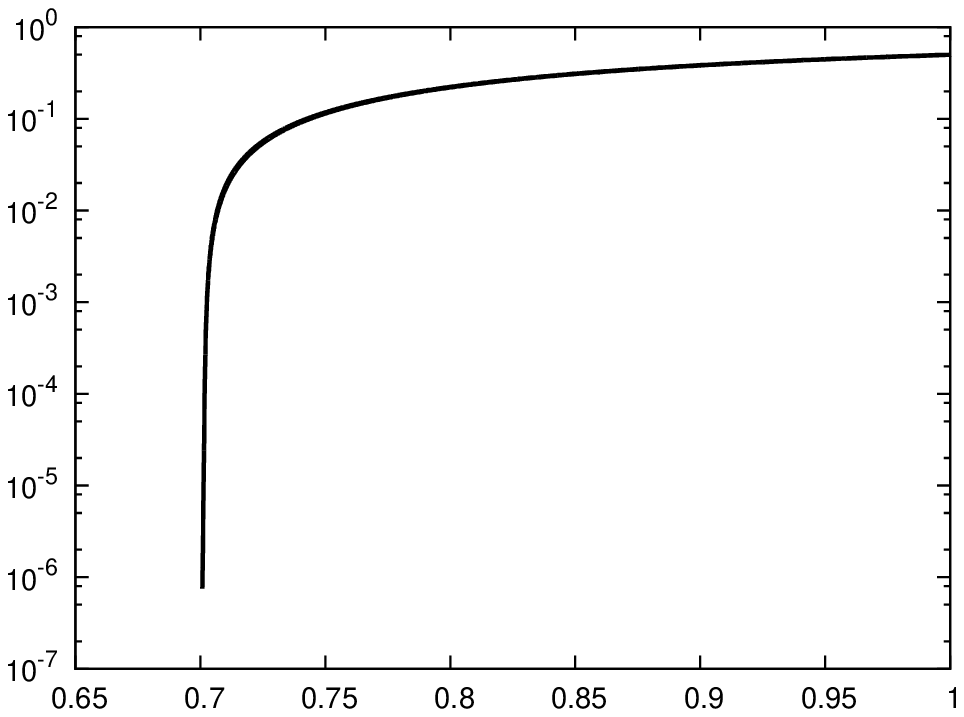,angle=270,width=4.7cm}
\psfig{file=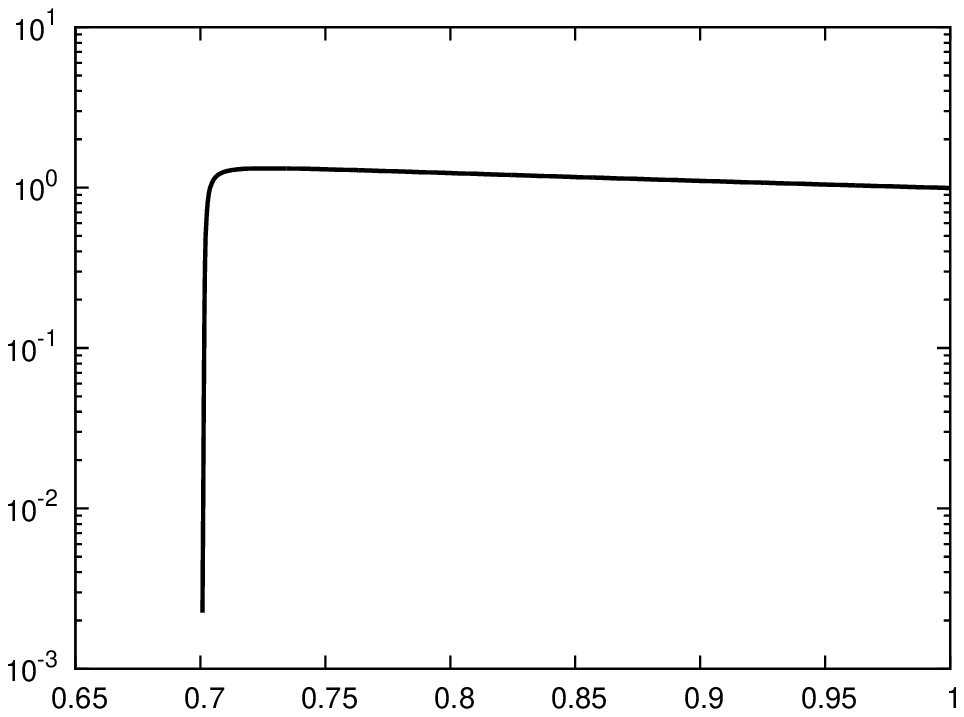,angle=270,width=4.7cm}}
\caption{The $1+\tilde g_2$ and the absolute magnitude of the beta function 
$\tilde\beta_2$ as the functions of the cutoff.}\label{massb}
\end{figure}

\begin{figure}
\begin{tabular}{cc}
\centerline{\psfig{file=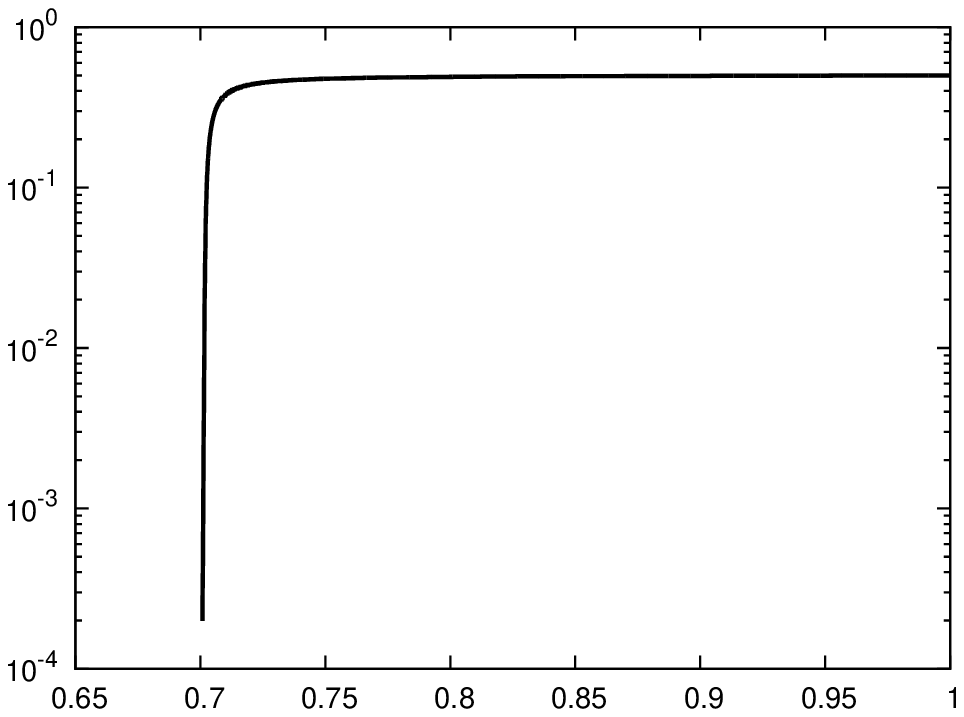,angle=270,width=4.7cm}
\psfig{file=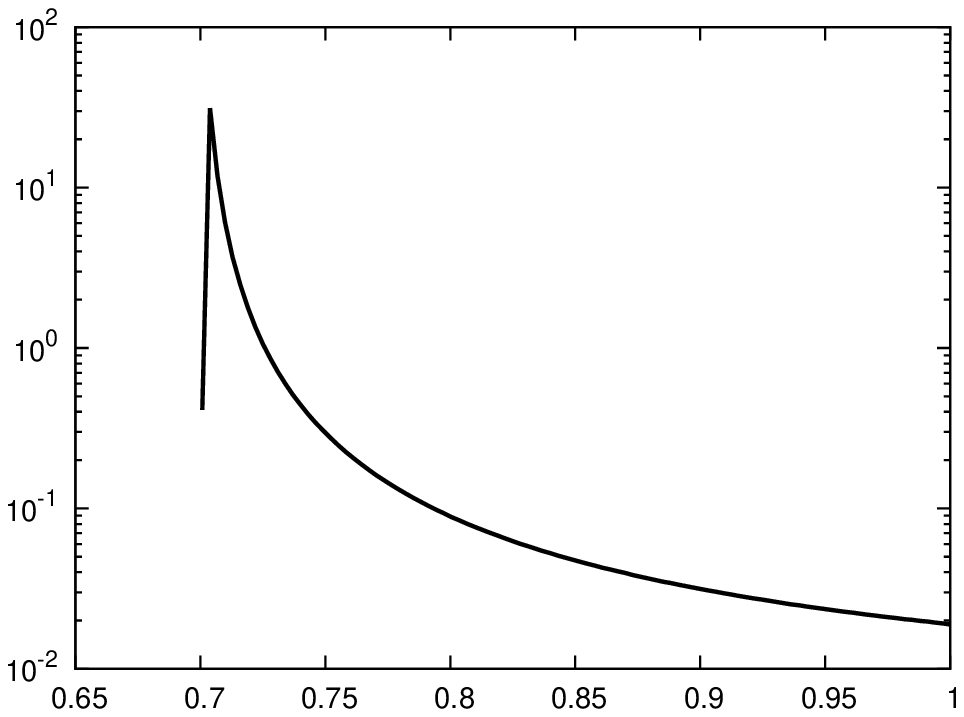,angle=270,width=4.7cm}}
\end{tabular}
\caption{The plot of $\tilde g_4$ (left) and its beta function (right).}\label{lambdab}
\end{figure}

\begin{figure}
\begin{tabular}{cc}
\centerline{\psfig{file=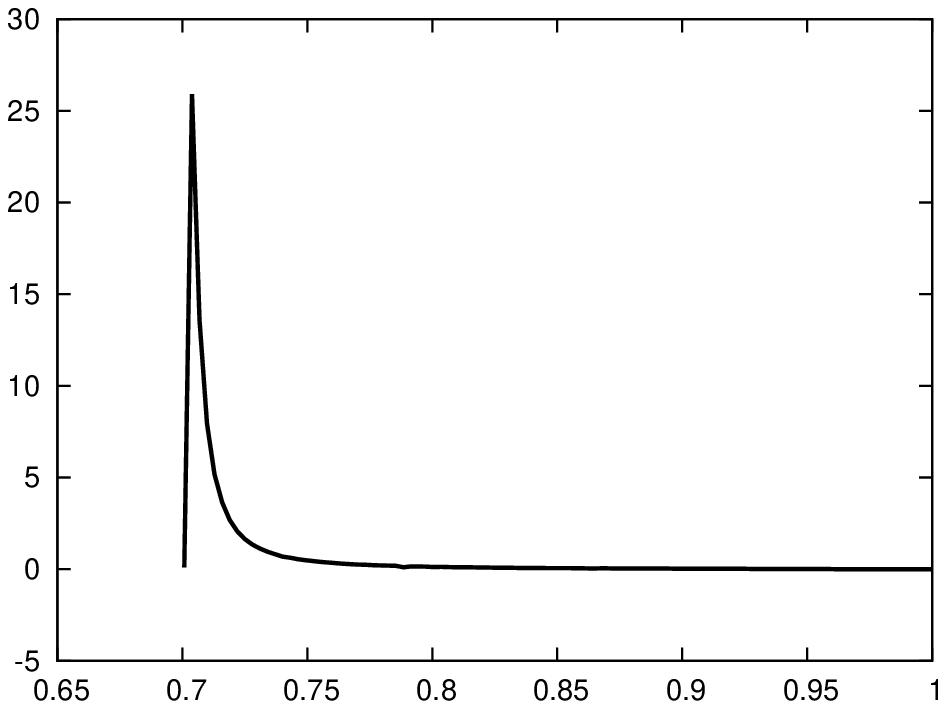,angle=270,width=4.7cm}
\psfig{file=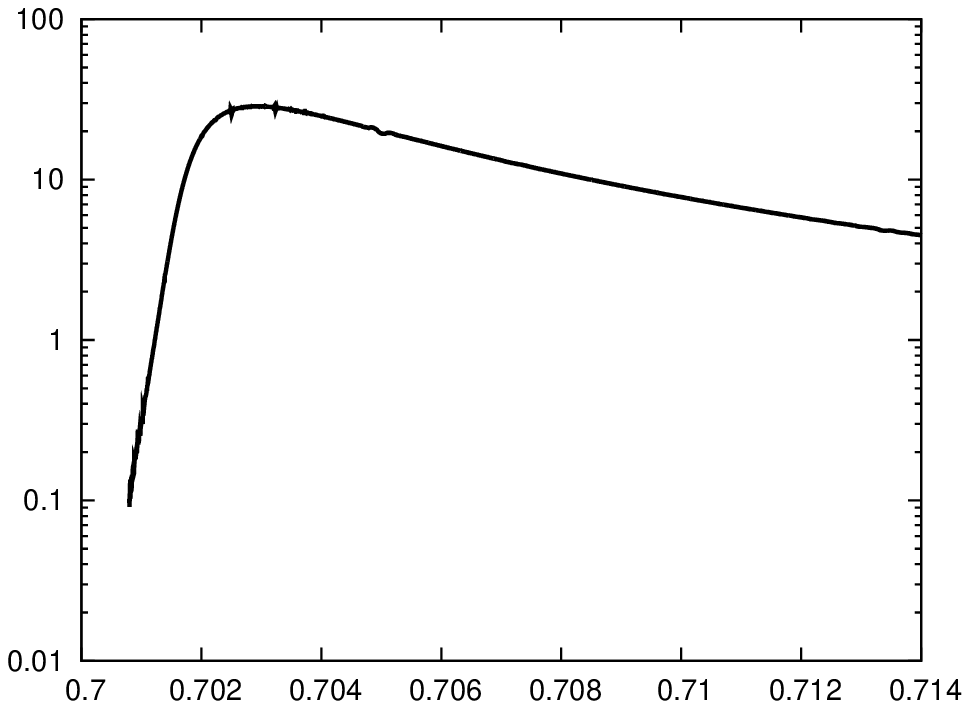,angle=270,width=4.7cm}} 
\end{tabular}
\caption{The same as Fig. \ref{lambdab} except for $\tilde g_6$ close to $k=k_{cr}$.}\label{higherlambdab}
\end{figure}

Let us now consider the numerical results in the symmetry broken phase, obtained for 
the bare parameters $g_{2B}=-0.5$, $g_{4B}=0.5$ and $g_{nB}=0$ for $n>4$. The initial 
value of the cutoff is chosen to be unity, $\Lambda=1$. 
The evolution of the first three dimensionless coupling constants at $\Phi=0$,
$\tilde g_n=k^{n-4}g_n$, together with their beta functions 
\be\label{dimlbetafc}
\tilde\beta_n=k^{n-4}\beta_n+(n-4)\tilde g_n
\ee
are shown in Figs. \ref{massb}-\ref{higherlambdab}. 
The dimensionless $\tilde g_2$ and its beta function, displayed on Fig. \ref{massb}
have a short scaling window at the vicinity of the initial cutoff when the strongly
irrelevant couplings die out followed by a relatively long, stable scaling regime
with $\tilde\beta_2\sim-1/k^2$. The dimensional mass square, $g_2$ is scale independent
in this scaling region which indicates that the theory is rather weakly coupled. Similar 
curves, shown for the fourth and sixth order coupling constant in Figs. \ref{lambdab}
and \ref{higherlambdab} show the long, stable scaling regime where $g_4$ is weakly,
$g_6$ is stronger irrelevant with little evolution.
The same features can be found sufficiently far from the critical point at the
other side, deeply inside of the the symmetric phase, too. There the stable scaling regime 
extends to the IR end point because interactions die out below the mass gap.
The differences between the deep symmetric and deep symmetry broken theories is
that this almost free scaling regime comes to an abrupt end, around $k_{cr}\approx0.7$
in these Figures.

The stable scaling regime ends with a strong interactions, despite the smallness of
the quartic coupling strength the higher order coupling constants sharply increase and
the potential looses the polynomial expandability. This is driven by the rapid building
up of the degeneracy, $1+\tilde g_2\to0$. The maximal speed of decrease, the peak in 
$\beta_4$, shown in the right plot of Fig. \ref{lambdab} locates $k_{cr}$ more precisely
at $k_{cr}\approx0.71$. The higher order coupling 
constants, for instance the evolution of $\tilde g_6$, shown in Fig. \ref{higherlambdab}
indicates the same crossover scale, except that the running coupling constants
pass a sharp peak before falling upon entering into the IR scaling regime.

\begin{figure}
\begin{tabular}{cc}
\centerline{\psfig{file=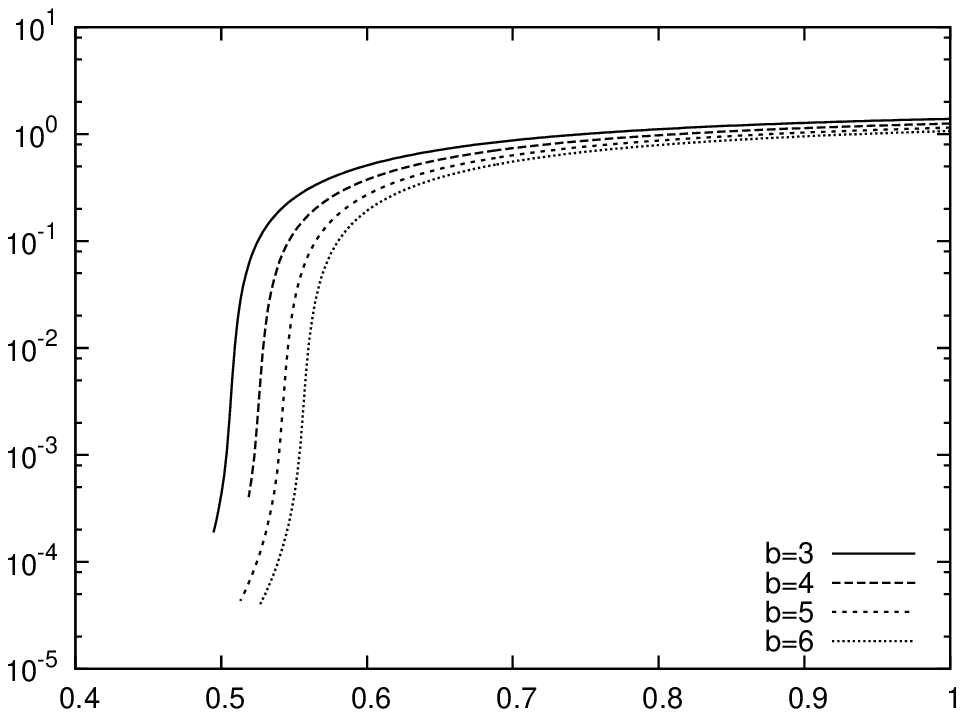,angle=270,width=4.7cm}
\psfig{file=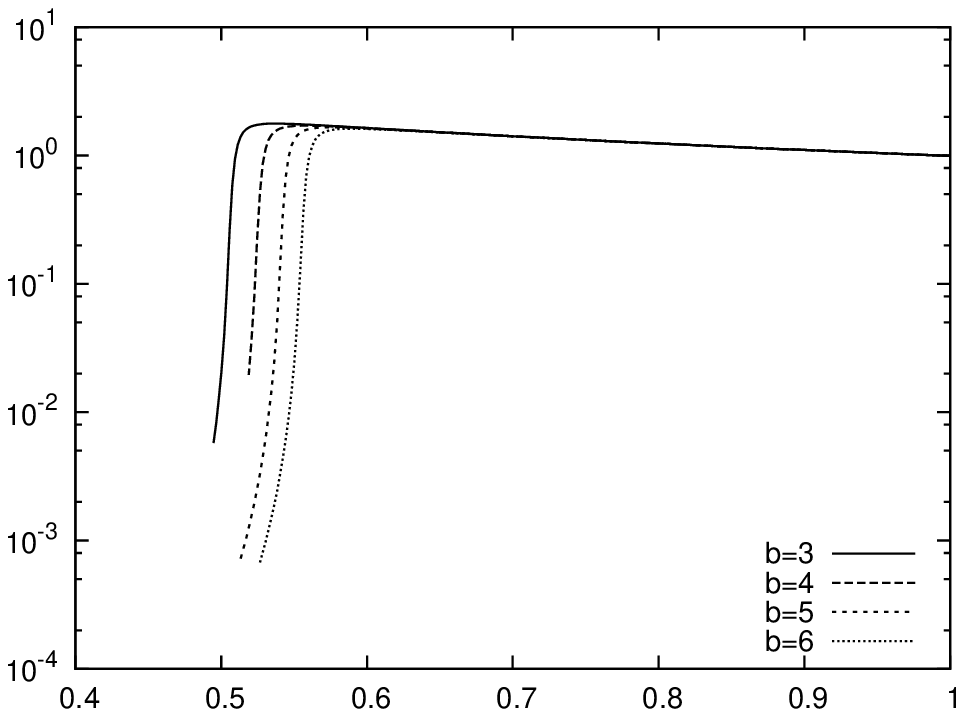,angle=270,width=4.7cm}} 
\end{tabular}
\caption{The evolution of $\tilde{k}^2+\tilde g_2$ and $\tilde\beta_2$ calculated from the 
effective action, cf. Eq. \eq{meas}.}\label{masse}
\end{figure}

\begin{figure}
\begin{tabular}{cc}
\centerline{\psfig{file=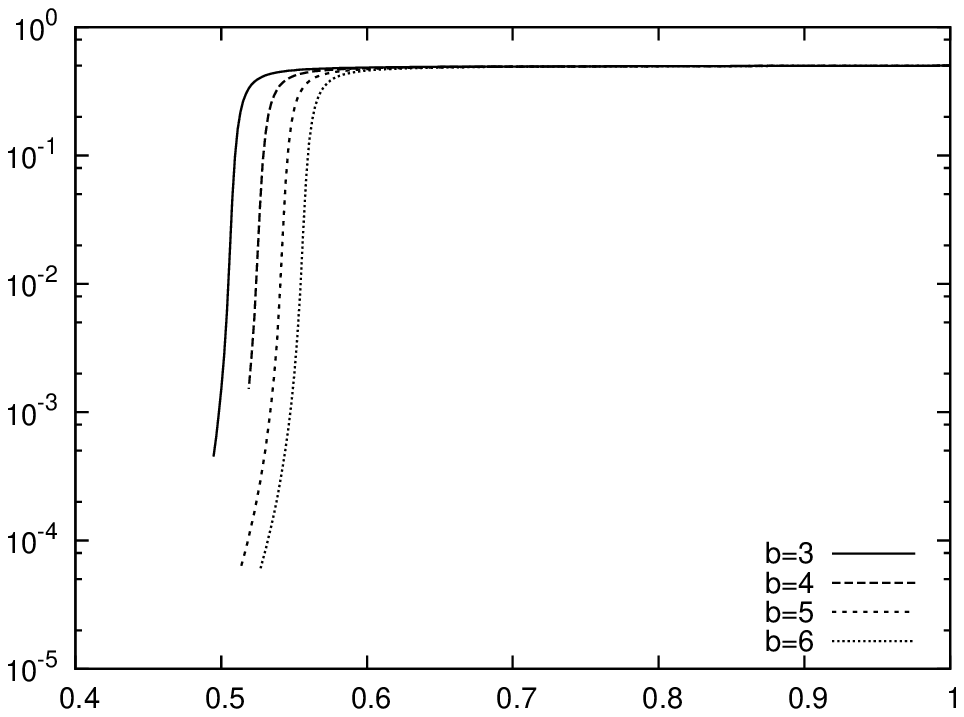,angle=270,width=4.7cm}
\psfig{file=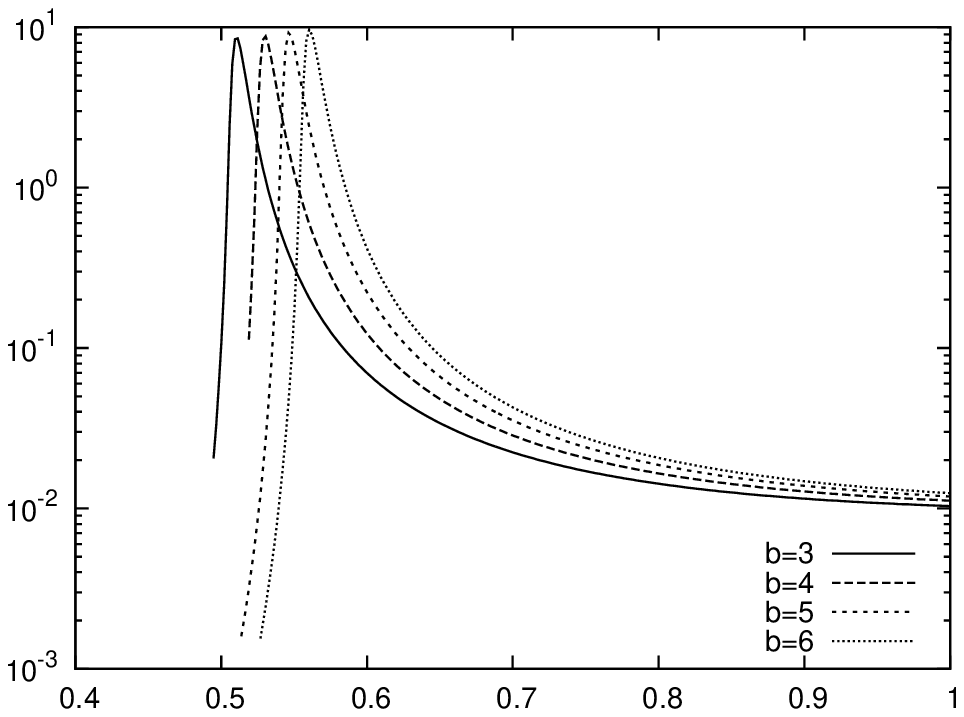,angle=270,width=4.7cm}} 
\end{tabular}
\caption{The plot of $\tilde g_4$ (left) and its beta function (right), calculated from the 
effective action.}\label{g4e}
\end{figure}

\begin{figure}
\begin{tabular}{cc}
\centerline{\psfig{file=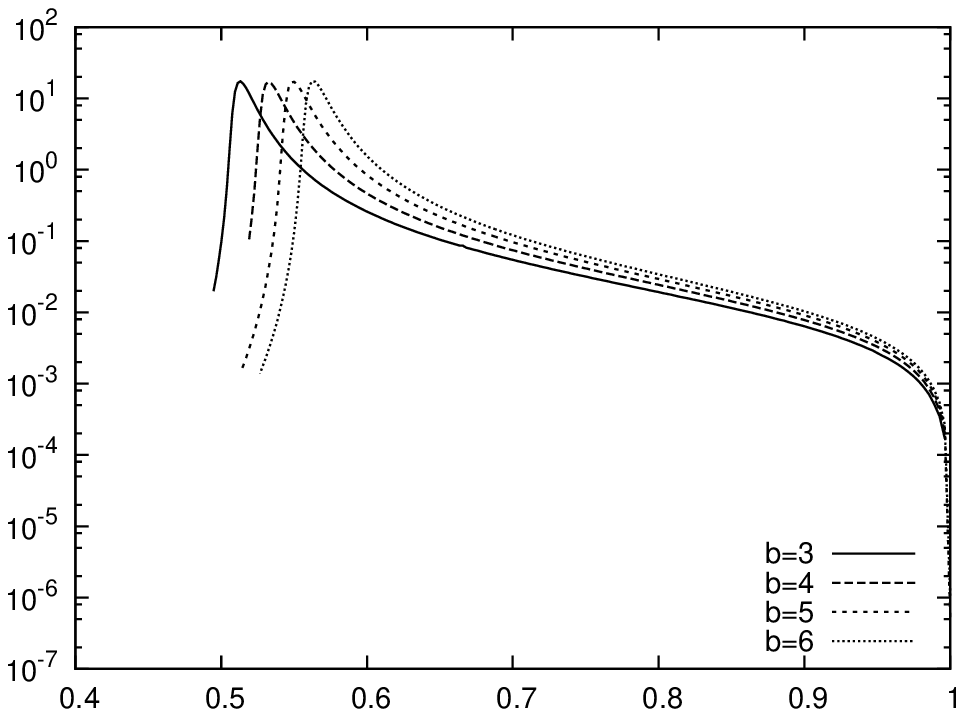,angle=270,width=4.7cm}
\psfig{file=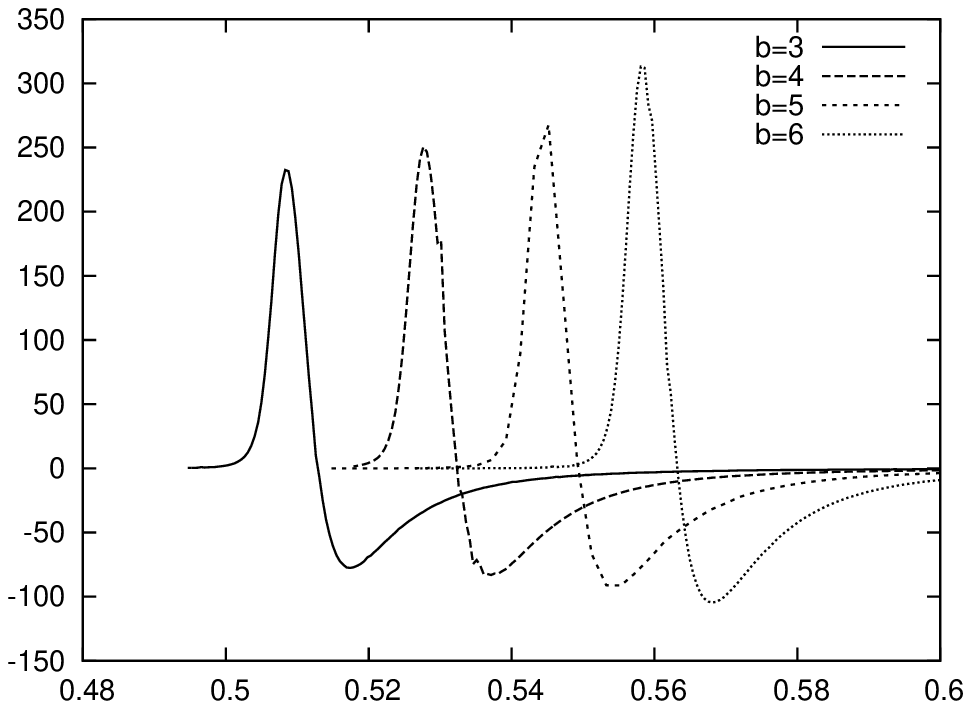,angle=270,width=4.7cm}}
\end{tabular}
\caption{The same as Fig. \ref{g4e} except for $\tilde g_6$.}\label{g6e}
\end{figure}

Qualitatively similar solution is found in solving Eq. \eq{smooth} for the effective 
potential. The renormalized trajectory of the same bare theory is depicted in
Figs. \ref{masse}-\ref{g6e}. A strongly degenerate theory is reached significantly 
later, after a longer weakly interacting scaling regime than in the same theory
treated by sharp cutoff as a result of a different set of irrelevant terms in the
sharp and smooth cutoffs. But apart of a global scale factor in $k$ the dimensionless 
quantities obtained by different regulators are similar in the stable scaling region.
The final approach to the degenerate theory is nonuniversal naturally.

\subsection{Condensation as rapid crossover}
How to interpret the precursor of condensation, a high degree of degeneracy of the action
before the IR end point, seen in these Figures? The answer can be given 
by inspecting the shape of the potential. The condensate of the scalar model is homogeneous, 
it consists of particles with vanishing momentum. Their density which is proportional to 
$\Phi^2$ can be controlled by an external, homogeneous source, coupled to the quantum field. 
The Maxwell-cut, observed at $k=0$, assures that the vacuum energy is independent of the 
density of the particles in the condensate in some density range, proportional to 
$\Phi^2_\mr{vac}$. Beyond this threshold density the repulsion becomes strong enough to 
place the additional particles at higher, non-condensed states. 

\begin{figure}
\begin{tabular}{cc}
\centerline{\psfig{file=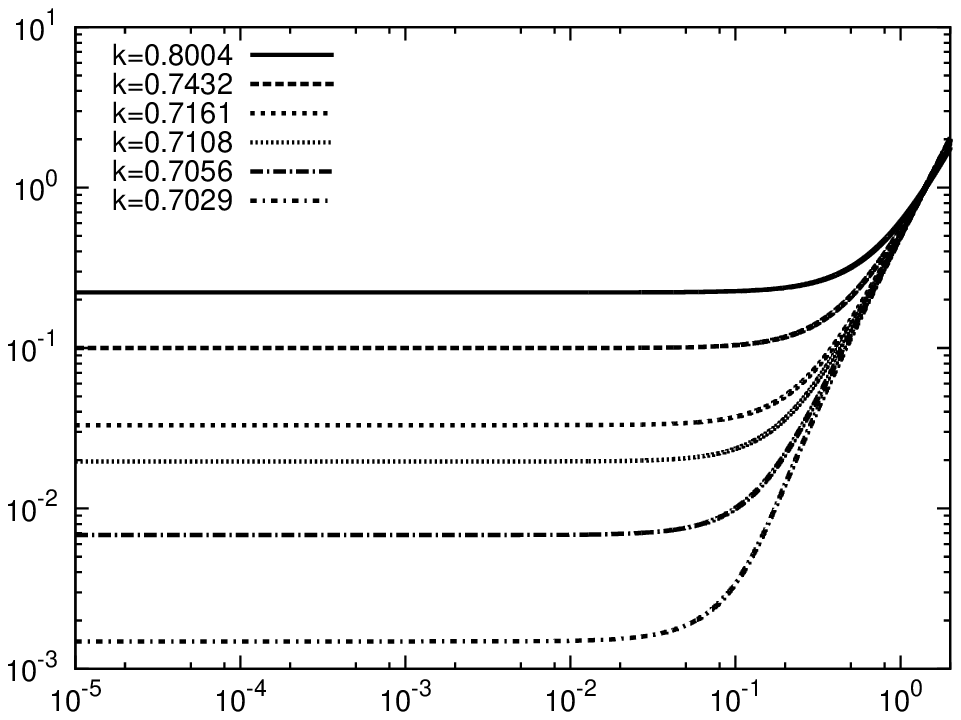,angle=270,width=4.7cm}
\psfig{file=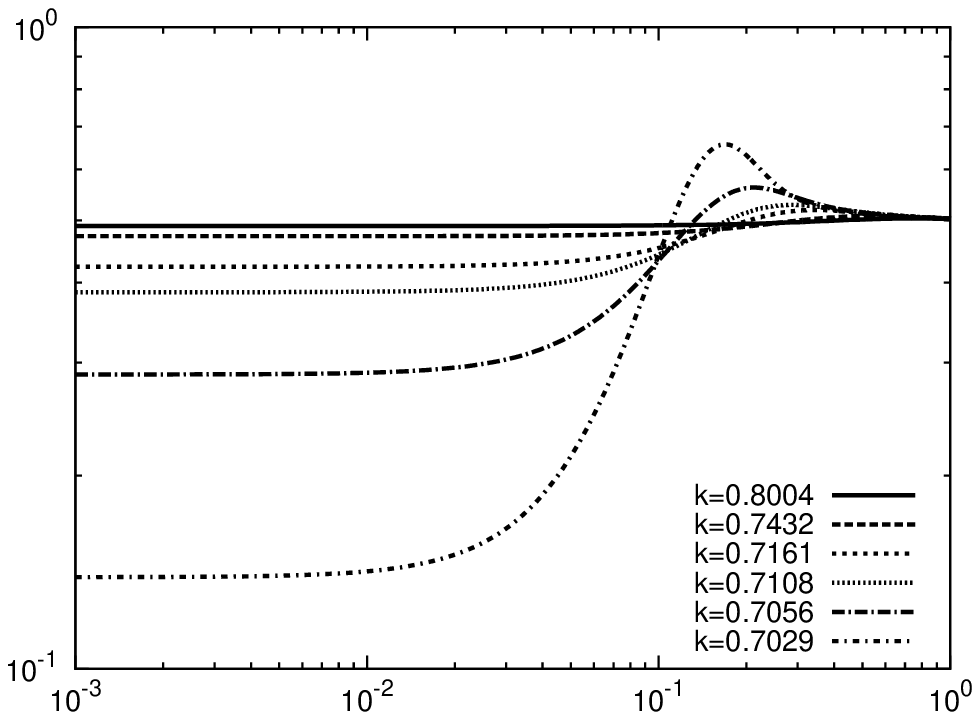,angle=270,width=4.7cm}}
\end{tabular}
\caption{The degeneracy, $1+\tilde g_2$, (left) and the quartic coupling strength
$g_4$ (right) as the function of the field $\Phi$.}\label{wh_deg_V2}
\end{figure}

This feature can be found at finite scale, too. Note that the action $S_k[\phi]$ is used
for the modes with scale $[k-\Delta k,k]$ only in the Wegner-Houghton scheme therefore
the bare potential captures the effective couplings of this local scale window. The curvature
and the fourth derivative of the bare potential, plotted in Figs. \ref{wh_deg_V2} for the
least few values of the cutoff before the program crashes show that the bare action becomes highly 
degenerate. This requires the potential $V^\mr{deg}_k(\Phi)=-k^2\Phi/2$ within the local potential 
approximation. Such a drastic change of the potential indicates a strong dressing 
provided by particles with wave vector $k=k_{cr}\approx0.73$ which make the energy almost 
degenerate in an interval of $\Phi$ and install a dynamical Maxwell-cut. We have a 
qualitatively similar degenerate plateau, bounded by an almost discontinuous jump when we 
consider the potential as the function of $\Phi^2$, a variable which is proportional to 
the condensate density. The derivative of this function, $dV(\Phi^2)/d\Phi^2$, is 
proportional to the binding energy. Thus the suppression of the binding energy,
infered from the left and the smallness of the two-body forces shown
on the right in Figs. \ref{wh_deg_V2} suggest that the interactions
are strongly suppressed in a dilute gas. As the degeneracy is approached the threshold
density until the free gas behaviour is observed increases.

\begin{figure}
\begin{tabular}{c}
\centerline{\psfig{file=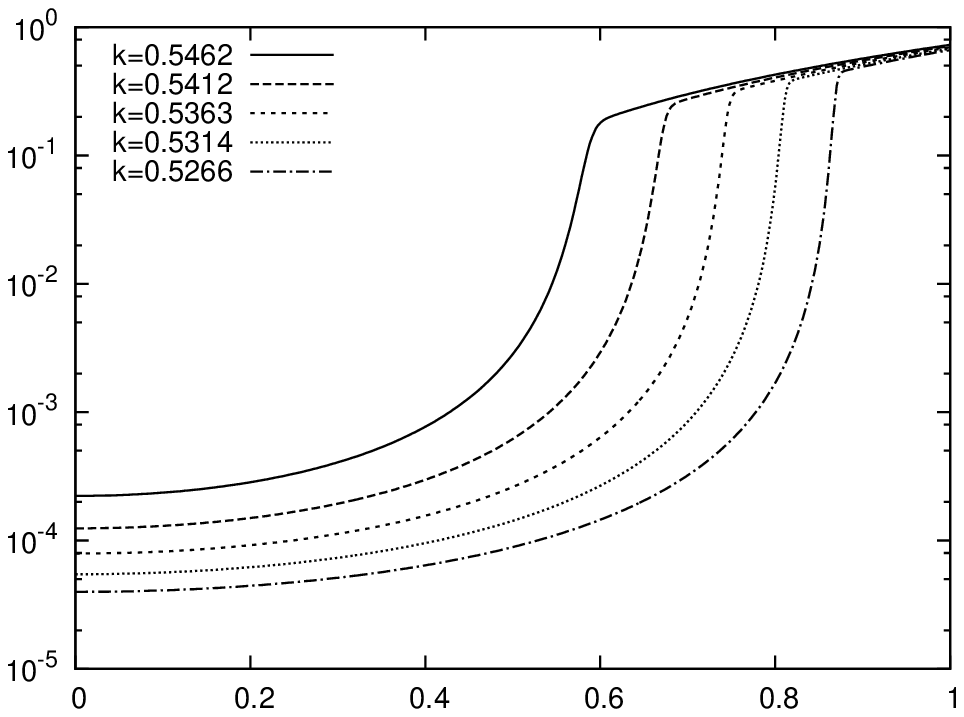,angle=270,width=4.7cm}}
\end{tabular}
\caption{The degeneracy, $\tilde{k}^2+\tilde g_2$ as the function of the field $\Phi$ of the effective action, 
cf. Eq. \eq{meas}.}\label{effa_deg_V2}
\end{figure}

Higher order coupling constants show a more involved scale dependence when
the cutoff is decreased. Though their dimensionless values drop in the UV scaling 
regime they display a peak at $k_{cr}$ which becomes taller with the increase of $n$. 
The large peak indicates that the particles interact with strong many-body forces at that scale. 
Thus the key difference between theories with and without condensate is that multi-particle
interactions reach a strong peak at finite scale for the latter and build new kind of
quasi-particles. The emergence of the simple but unusual potential $V^\mr{deg}_k(\Phi)$, 
the dynamical Maxwell-cut, preceded by the dramatic raise and fall of the 
higher order interaction strengths indicates the emergence of radically different 
quasiparticles than those in the perturbative, ``empty`` vacuum. This difference shows 
up in their dressing, their structure in terms of the bare particles though their 
dispersion relation remains usual, being controlled by the unbroken space-time symmetries.

\subsection{Quantum censorship}
We are finally in the position to address the questions raised by the tree-level evolution 
which predicts a strongly degenerate action at finite scale. Is this result confirmed by our numerical
solution where the quantum fluctuations are properly taken into account? Is the strategy,
based on the integration of the evolution equation protected against nonintegrable
singularities? The undeniable fact is that our algorithm stops at finite scale. 
The linear algebra, used in determining the evolving coefficients of the spline functions, 
becomes singular for degenerate action when the right hand side of the evolution equation 
\eq{wh} is vanishing. The program optimizes the step size, $\Delta k$, automatically and stops 
when the irregularities in the interpolation of the derivatives or in the integration of the 
differential equation are significantly larger than the precision of the number representation 
in the machine. Such a dynamically adjusted step size is very small before stopping to keep
the small parameter $\epsilon_k$ under control.

One can imagine the following two scenarios in the presence of unlimited computer power 
when the numerical accuracy can be increased without limit. 
\begin{enumerate}
\item The action does not become exactly degenerate, the dynamical readjustment of the step 
size $\Delta k$ in the numerical algorithm is sufficient to resolve the variation of the 
action and the program continues functioning down to the IR end point. 

\item Whatever computer power we employ, the algorithm stops at some finite  scale
because $\Delta k$ shrinks too fast. 
\end{enumerate}

In the first scenario the scale $k_{cr}$ does not exists, $k_{tr}=0$ but the loop 
corrections contrive the dynamical Maxwell-cut which used to be explained in terms 
of semiclassical physics. This mechanism, the prevention of reaching a singular dynamics
by the help of quantum fluctuations, is called Quantum Censorship \cite{sg}.
In the second case the scale where the algorithm stops during increased precision
converges (being a decreasing, positive series) and it sets $k_{tr}$ where
the derivation of the evolution equation for the bare action is lost but that for 
the effective action remains valid and only its ansatz becomes unjustifiable. 
The sudden drop of the beta functions when the scale is lowered after their peak is 
reached at $k_{cr}$ suggests that the evolution equation may remain integrable at the 
phase boundary of the incomplete theories at this scale. 

\begin{figure}
\begin{tabular}{cc}
\centerline{\psfig{file=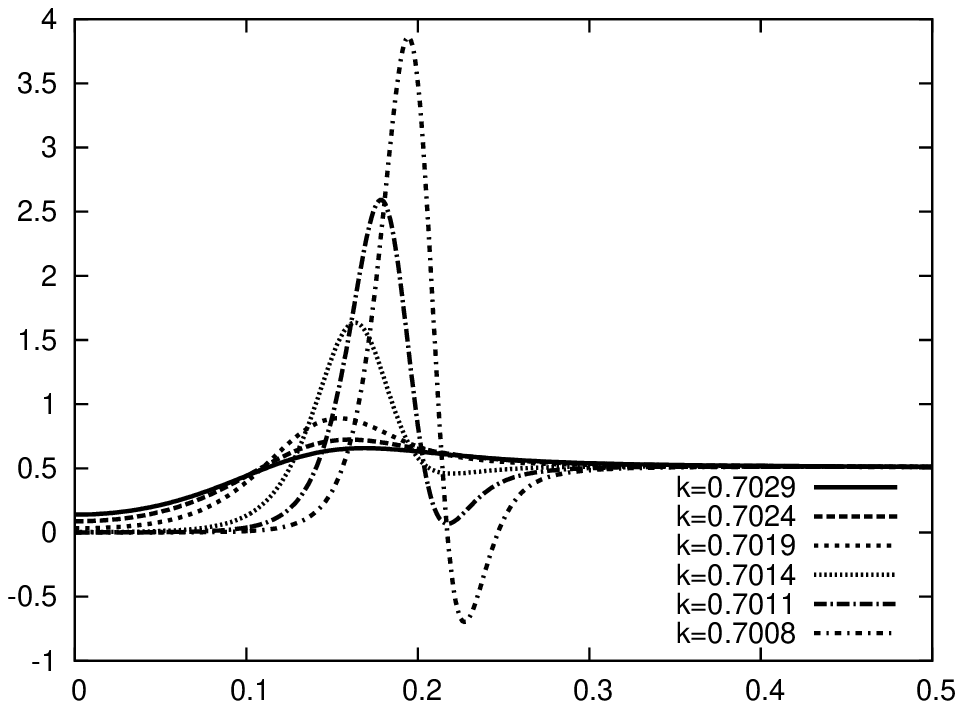,angle=270,width=4.7cm}
\psfig{file=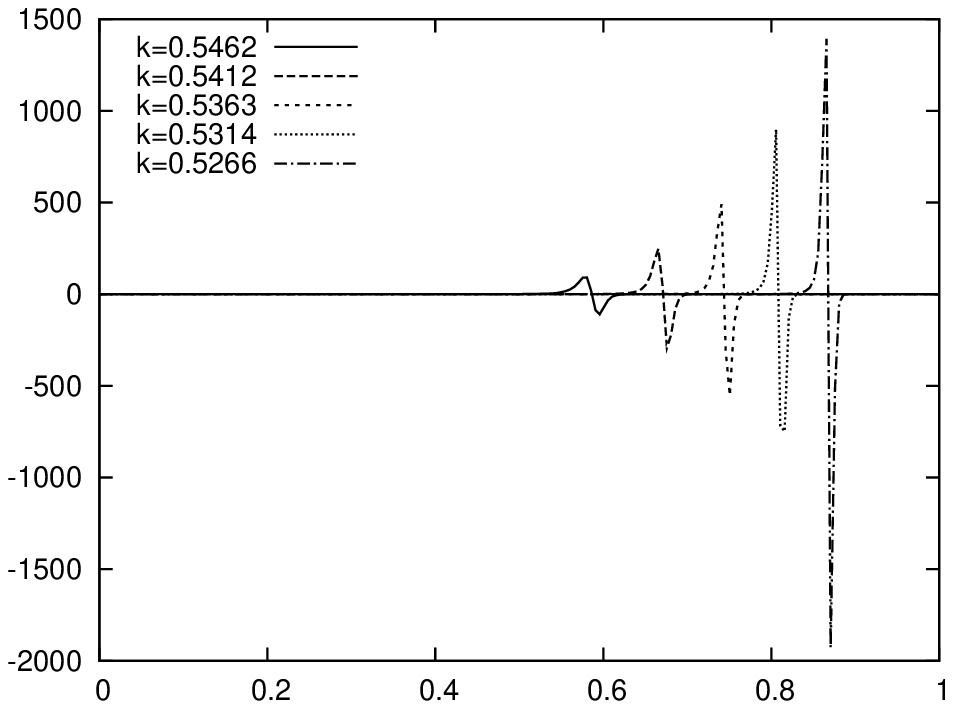,angle=270,width=4.7cm}} 
\end{tabular}
\caption{left: The quartic coupling strength, $g_4$, of the bare potential shown as 
the function of the field $\Phi$. The function $g_4(\Phi)$ which decreases with the cutoff around $\Phi=0$
is displayed only for the last few steps of the algorithm for the bare (left) and the effective
(right) potential.}\label{wh_V4_deepkink}
\end{figure}

The seemingly uninterrupted and sudden decrease of the degeneracy in the first plots of 
Figs. \ref{massb} and \ref{masse} taking place places far from $k=0$ makes it more likely 
that scenario 2. is realized in this scalar model, contrary to the periodic model of Ref.
\cite{sg}. Though the eventual stabilization beyond 
the accuracy of our solution we can point to the likely source of the stop of the algorithm. 
The fourth derivative of the potential is shown in Fig. \ref{wh_V4_deepkink} for the last 
few steps before crashing the program. The cutoff is decreased gently and the eight order 
splines display, a regular looking potential in this final stage of the evolution except 
at the edge of the degeneracy. One observes a shock wave squeezed against the beginning
of field values where the action is non-degenerate and the evolution is regular.
Thus the final value of $k_{tr}$ seems to depend on the way this shock wave is ultimately 
resolved.

\section{Summary}\label{sum}
Any attempt to evaluate functional integrals is related to some kind of expansion. This
strategy is seriously hampered when the integrand is a constant, the action is degenerate.
The most natural setting of this problem, the emergence of a condensate in a vacuum with 
a spontaneously broken symmetry, is considered in this work. Our method is the numerical 
construction of the renormalization of the local potential of the $\phi^4$ model
without assuming any ansatz for the potential. 

It is found that the weakly coupled evolution of the local potential is interrupted
by a sudden building up of degeneracy both of the bare and the effective action at finite
scale. This degeneracy, a dynamical Maxwell-cut, is interpreted as a precursor of the 
condensate and the usual Maxwell-cut of the true vacuum.

The rather abrupt appearance of the degeneracy along the renormalized trajectories poses a wonderful
problem both on the analytical and numerical sides. On the analytical side it is not
clear how to handle a degenerate action and whether the evolution equation remains
integrable. The challenge of the numerical treatment of an almost degenerate action is 
to overcome the errors arising from the limited precision of number representation in the 
computer. As a result, we had to be satisfied with a conjecture only, namely that the
loop contribution seemed to be inefficient to prevent the degeneracy and the evolution
equations might become singular.

We followed the evolution of the local potential ansatz. An obvious improvement 
is to consider the higher orders of the gradient expansion. We hope to report soon results
corresponding to the solution of the evolution equation by including the wave function 
renormalization.

\appendix
\section{The numerical algorithm}\label{algorithm}
Let us write the evolution equation \eq{wh} as 
\be\label{sdwh}
\partial_ku=-\frac{k^3}{16\pi^2}r'
\ee
where $v_k(\phi)=V_k''(\phi)$ and $r_k(\phi)=v_k'(\phi)/[k^2+v_k(\phi)]$. The boundary
conditions (i) $v_k(0)=0$, (ii) $v_k(\Phi_\mr{max})=f_k$ and (iii) $v_\Lambda(\phi)=v_B(\phi)$ 
with given functions $f_k$ and $v_B(\phi)$ are sufficient to have a unique solution.

The function $v_k(\phi)$ is given in a spline form \cite{berzins,consoli}, represented as the sum of 
eight order Chebyshev polynomials of the field $\phi$ in order to optimize the convergence
of the solution when the order of the spline polynomial or their number is increased
\cite{schryer}. The partial differential equation \eq{sdwh} is written as a set of 
ordinary differential equation for the coefficients, considered as functions of $k$. 
The error of such a numerical algorithm has been studied in Ref. \cite{berzin}.
The program stops in our implementation when the interpolation of the derivatives 
of the potential with respect to $k$ or $\phi$ is unreliable. Since the approximate
form of these derivatives is a ratio involving the potential at different points
which becomes $0/0$ in the degenerate limit the degeneracy is significantly
larger than the number representation accuracy when the algorithm stops.
The boundary conditions (i) and (iii) are easy to implement,
(ii) can be replaced by an algebraic equation, written as
\be\label{boundc}
\beta_k(\Phi_\mr{max})r_k(\Phi_\mr{max})=\gamma_k(\Phi_\mr{max},v_k(\Phi_\mr{max}),v'_k(\Phi_\mr{max}),
V_k(\Phi_\mr{max}),\partial_kV_k(\Phi_\mr{max}))
\ee
in the functional notation in terms of given functions $\beta$ and $\gamma$.

The algebraic equations for the coefficients of the Chebyshev polynomials remain well
defined in the limit $\beta,\gamma\to0$. The boundary condition \eq{boundc} is imposed
in this null-equation limit. The solution is carefully monitored in and was found stable
as the upper limit of the equation was by sent to infinity, $\Phi_\mr{max}\to\infty$. 
The potential $U_k(\phi)$ was found to be convergent and $\Phi_\mr{max}$
was finally fixed by having relative error $10^{-6}$ up to field values
higher than $\Lambda$.

The algorithm has been tested by the choice $v_k(\phi)=V_k'(\phi)$ but it was found to be
less stable and more time consuming.

\section{Wilson-Fisher fixed point}\label{wilsfis}
The three dimensional scalar model displays a nontrivial fixed point at $d=4-\epsilon$
dimensions which is nonperturbative for $\epsilon\approx1$. Our numerical algorithm to 
integrate the evolution equation Eq. \eq{smooth} is checked by the recovery of this
fixed point. The choice of the suppression term \cite{litim}
\be\label{optreg}
R_k(p^2)=(k^2-p^2)\Theta(k^2-p^2)
\ee
leads to a simple differential equation for the fixed point potential 
\be\label{fpeq}
0=-dV+\frac{d-2}2\Phi V'+\frac{4\Gamma(d/2)}{2^{d+1}\pi^{d/2}d}
\frac{k^d\Phi^3}{\Phi^3+(k^{-2}\Phi^2-k^{d-4})V'+k^{d-4}\Phi V''}.
\ee
By assuming the polynomial ansatz
\be\label{fixp}
V(\Phi)=k^d\sum\limits_{n=0}^\infty\frac{\lambda_n}{n!}\left(\hf\Phi^2k^{2-d}-\lambda_1\right)^n
\ee
for the solution the insertion of this form into the fixed point equation \eq{fixp} 
gives the expansion coefficients $\lambda_n$, $n\ne1$ in terms of $\lambda_1$ which is
related to the position of the minimum, $\Phi_{min}=\sqrt{2\lambda_1}k^{(d-2)/2}$.

\begin{figure}
\begin{tabular}{c}
\centerline{\psfig{file=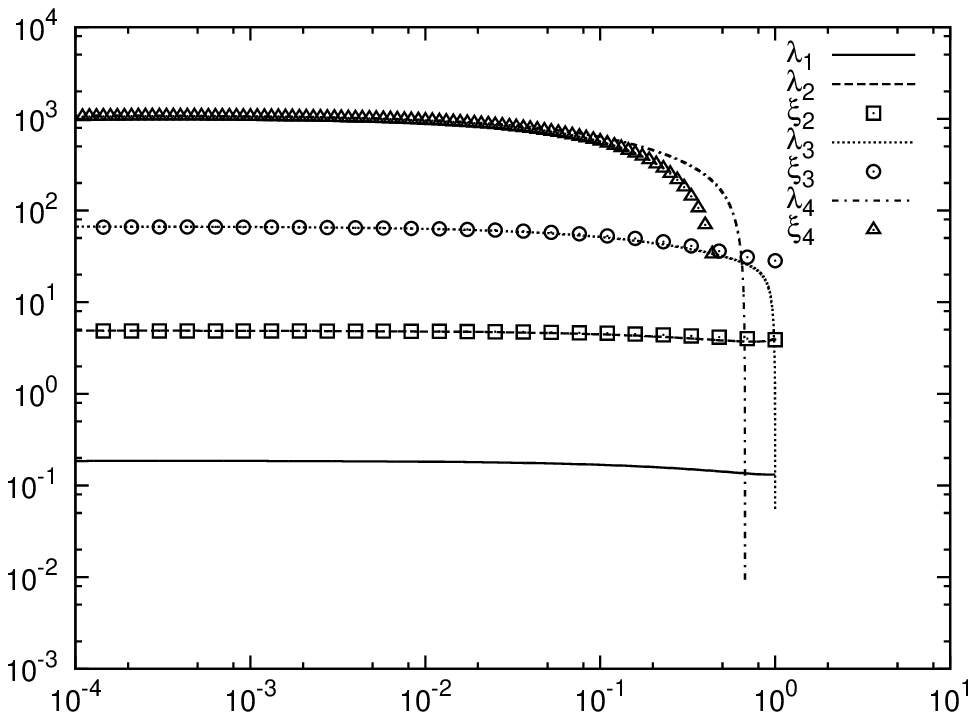,angle=270,width=4.7cm}} 
\end{tabular}
\caption{Approach of the Wilson-Fisher fixed point in $d=3$.}\label{wffp}
\end{figure}

The numerical solution of the evolution equation \eq{smooth} with the regulator \eq{optreg}
in $d=3$ dimensions show nicely the fixed point \eq{fpeq}. This can be checked by following
the minimum of the effective potential $\Phi_{min}$ and reconstructing the coefficients
$\lambda_n$ for $n\ne1$ from $\lambda_1=\Phi_{min}^2k^{2-d}/2$. The expressions $\xi_n$
obtained in this manner approach to $\lambda_n$ as long as the trajectory is running into
the fixed point as seen in Fig. \ref{wffp}, obtained for the bare mass square
$g_2=m^2=-0.131058371$, and coupling strength $g_4=4$, set at $\Lambda=1$. The numerical integration
stopped at $k=0.0001$. All digits are significant in the mass, without such a fine tuning
the trajectories leave the repelling fixed point region earlier.

\section*{Acknowledgements}

S. Nagy, K. Sailer and J. Polonyi acknowledge an MTA-CNRS grant.

\end{document}